\documentclass[11pt,oneside]{article}

\usepackage[utf8]{inputenc}
\usepackage[T1]{fontenc}
\usepackage[margin=1in]{geometry}
\usepackage{microtype}
\usepackage{amsmath,amssymb}   
\usepackage{newtxtext,newtxmath}

\usepackage{graphicx}
\setkeys{Gin}{width=\linewidth,keepaspectratio}
\usepackage{booktabs}
\usepackage{tabularx}
\usepackage{makecell}
\usepackage{xcolor}
\usepackage{algorithm}
\usepackage{algpseudocode}
\usepackage{float}                 
\usepackage[section]{placeins}     
\usepackage{appendix}

\usepackage{caption}
\usepackage{url}
\usepackage[numbers,sort&compress]{natbib}
\usepackage[hidelinks]{hyperref}
\usepackage{xurl}                  

\usepackage{listings}
\usepackage{listingsutf8}
\lstdefinestyle{cleanTeX}{
  basicstyle=\ttfamily\footnotesize,
  breaklines=true,
  breakatwhitespace=false,
  breakindent=0pt,
  postbreak=\mbox{\textcolor{gray}{$\hookrightarrow$}\space},
  frame=single,
  backgroundcolor=\color{gray!5},
  keywordstyle=\color{blue!70!black},
  commentstyle=\color{gray!80},
  showstringspaces=false,
  columns=fullflexible,
  keepspaces=true,
  upquote=true,
  aboveskip=6pt,
  belowskip=6pt
}
\lstdefinelanguage{JavaScript}{
  morekeywords={break,case,catch,continue,debugger,default,delete,do,else,
    finally,for,function,if,in,instanceof,new,return,switch,this,throw,try,
    typeof,var,let,const,void,while,with,yield,class,enum,export,extends,
    import,super,implements,interface,private,public,protected,static,await,async,get,set},
  sensitive=true,
  morecomment=[l]{//},
  morecomment=[s]{/*}{*/},
  morestring=[b]',
  morestring=[b]"
}
\lstdefinelanguage{TypeScript}[]{JavaScript}{
  morekeywords={type,namespace,abstract,readonly}
}

\usepackage{pifont}
\newcommand{\cmark}{\ding{51}}  
\newcommand{\xmark}{\ding{55}}  
\newcommand{\warn}{\ding{115}}  
\DeclareUnicodeCharacter{2713}{\cmark}
\DeclareUnicodeCharacter{2717}{\xmark}
\DeclareUnicodeCharacter{26A0}{\warn}

\graphicspath{{./}{latex/}}

\title{From Paper to Structured JSON: An Agentic AI Workflow for Compliant BMR Digital Transformation}

\author{
Bhavik Agarwal\thanks{ Equal contribution.},
Nidhi Bendre\footnotemark[1],
Viktoria Rojkova \\
MasterControl AI Research \\
\texttt{\{bagarwal,nbendre,vrojkova\}@mastercontrol.com}
}

\date{}

\begin{document}
\maketitle

\begin{abstract}
Pharmaceutical manufacturers generate thousands of batch manufacturing records (BMRs) annually. Under FDA 21 CFR Part 211 and EU GMP guidelines, these 100+ page documents mix tables, calculations, images, and handwritten annotations and must be retained for decades \citep{fda2024cfr211,ema2022gmp}. Current digitization approaches sit between two extremes: generic document extraction tools that lack pharmaceutical domain understanding, and industry-specific systems that assume standardized digital inputs \citep{aizon2025batch,llamaextract2024}. As a result, manual conversion remains the norm, requiring hours of quality review per document and scaling linearly with volume \citep{pharmatech2024bmr}.

We present an agentic AI workflow that transforms unstructured BMRs into compliant, structured JSON through parallel processing and multi-layer validation. The system uses token-based chunking to handle long documents, processes chunks in parallel with schema-guided LLM extraction, and merges results while preserving the hierarchical Group--Phase--Step structure. A TypeScript-like schema with 11 content types (tables, calculations, images, data forms, etc.) constrains extraction so that pharmaceutical semantics are preserved.

The architecture implements three validation layers: (i) syntactic validation of JSON and tag structure, (ii) structural validation of class separation and referential integrity, and (iii) pharmaceutical compliance checks aligned with GMP expectations. We introduce coverage metrics that quantify extraction quality, including crude and context-aware word coverage and preservation of tables, images, and calculations.

On three representative real-world BMRs (15--66 pages), our system achieves composite confidence scores between 82.08\% and 89.13\% while maintaining perfect hierarchy, sequence, and cross-reference preservation and near-perfect fidelity for calculations, conditional logic, and units. Compared to manual review baselines of several hours per document, the parallel architecture reduces processing time to minutes or tens of minutes on standard infrastructure (single GPU, up to 8 parallel workers). Remaining challenges include OCR noise on historical documents (especially handwriting) and occasional cross-chunk context loss in very long ($>150$-page) records.

Overall, parallel schema-guided extraction with pharmaceutical-specific validation enables practical BMR digitization at scale, unlocking decades of manufacturing data for analysis while keeping humans in the loop for quality decisions that affect patient safety.
\end{abstract}

\section{Introduction}
\label{sec:intro}

The pharmaceutical manufacturing sector operates within a strict regulatory framework that requires comprehensive documentation throughout the production life cycle. Batch Manufacturing Records (BMRs), as mandated by regulatory authorities, including the United States Food and Drug Administration (21 CFR Part 211) and the European Medicines Agency (EU GMP Guidelines), constitute critical quality system components that document the complete manufacturing history of pharmaceutical products \citep{fda2024cfr211, ema2022gmp}. These records encompass raw material specifications, equipment utilization, process parameters, quality control results, deviation documentation, and corrective action implementation. BMRs serve multiple critical functions: enabling traceability for regulatory audits, facilitating product recall procedures, and establishing fundamental safeguards for patient safety \citep{pharmatech2024bmr}.

\subsection{Current State of Documentation Practices}

Contemporary pharmaceutical manufacturing facilities exhibit a stark contrast between sophisticated production technologies and outdated documentation methodologies. Despite significant technological advancement in manufacturing processes, documentation practices remain predominantly paper-based and manual in nature \citep{ispe2023digitization}. Production operators continue to manually transcribe readings from digital instrumentation onto paper-based forms, perform mathematical calculations without computational assistance, and collect physical signatures. Quality assurance personnel subsequently have to manually review these documents, which often exceed 150 pages per batch, before they can be archived in physical storage facilities. Retrieval of documentation for regulatory investigations or quality audits requires physical access to these archived materials followed by manual examination.

This reliance on paper-based documentation systems represents a fundamental operational paradox. Modern pharmaceutical organizations demonstrate capabilities in precision drug engineering while employing documentation methodologies that have remained largely unchanged since the 1980s \citep{mckinsey2025genai}.

\subsection{Previous Research and Existing Approaches}

The literature describes two different technological approaches developed to tackle the challenges of pharmaceutical documentation, but successfully integrating them has proven difficult. Newer artificial intelligence technologies have demonstrated capabilities in document extraction and digitization. Commercial solutions such as LlamaExtract \citep{llamaextract2024} and integrated frameworks combining LangChain with Pydantic validation have achieved success in converting standard business documentation---including invoices, contracts, and financial reports---into structured JSON formats \citep{langchain2024}. However, these general-purpose solutions show significant limitations when applied to pharmaceutical documentation, particularly in their inability to accommodate domain-specific requirements.

On the other hand, the pharmaceutical sector has developed specialized digitization platforms tailored to industry-specific requirements. Organizations such as Aizon have implemented intelligent batch record systems that automate manufacturing workflows and facilitate real-time data capture \citep{aizon2025batch}. Electronic Batch Record (EBR) systems have incorporated review-by-exception algorithms that reduce quality review time for newly manufactured batches \citep{mastercontrol2023ebr}. Nevertheless, these pharmaceutical-specific solutions primarily address prospective data capture rather than document conversion. Their reliance on standardized input formats makes them unsuitable for processing the diverse document formats accumulated over decades of manufacturing operations.

BMR digitization initiatives currently depend on specialized service organizations that employ manual data entry personnel for document conversion. While this methodology ensures accuracy and regulatory compliance, the associated economic costs and mundane requirements make it impractical for large-scale digitization programs \citep{deloitte2023digitization}.

\subsection{The Critical Gap}

The existing technological landscape reveals a critical gap at the intersection of AI and the pharmaceutical domain. Current AI-based extraction tools, while having sophisticated document processing capabilities, lack the pharmaceutical domain knowledge necessary to ensure regulatory compliance and maintain data integrity. Conversely, pharmaceutical-specific systems, while incorporating requisite compliance features, lack the flexibility necessary to accommodate the natural variation present in documents. This technological gap leads to large amounts of manufacturing data, which holds decades of process knowledge, being difficult to use for analysis or process improvement \citep{nature2024pharmaai}.

No existing solution successfully integrates large language model (LLM) capabilities with pharmaceutical domain expertise to process BMR documents while maintaining regulatory compliance standards. This limitation constitutes a fundamental barrier preventing pharmaceutical organizations from leveraging manufacturing intelligence for continuous improvement and process optimization initiatives.

\section{Problem}

\subsection{Operational Inefficiencies and Compliance Challenges}

Manual BMR processing workflows represent an operational burden within pharmaceutical manufacturing operations. Empirical analysis demonstrates that each batch record requires approximately three hours of Quality Assurance review time \citep{ispe2023metrics}. A representative mid-scale manufacturing facility producing 100 batches monthly generates approximately 1{,}200 BMRs annually, accumulating to tens of thousands of documents over a decadal period. Time-motion studies indicate that production operators allocate approximately 30\% of available working time to documentation activities rather than value-generating manufacturing operations \citep{lean2024pharma}. These inefficiencies show as delayed batch release cycles, increased inventory costs, and operational disruptions during regulatory inspections.

Optical character recognition (OCR) and document extraction technologies demonstrate inadequate performance when applied to pharmaceutical batch records. While these systems successfully extract textual content, they lack capabilities for validating mathematical calculations, finding missing records for fixing problems, or maintaining the comprehensive audit trails mandated by 21 CFR Part 11 electronic records requirements \citep{fda2024part11}. These technological limitations introduce regulatory compliance risks that may precipitate manufacturing line suspensions and necessitate resource-intensive remediation programs.

\subsection{Documentation Errors and Quality Implications}

Peer-reviewed studies in pharmaceutical quality literature identify documentation errors as a predominant source of manufacturing deviations, consistently ranking it among the primary factors in quality incidents \citep{pqri2023deviations}. Manual documentation processes introduce multiple errors: operators may switch numerical values during equipment reading transcription, quality reviewers may fail to identify missing authorization signatures, and calculations may contain undetected computational errors. These documentation deficiencies extend beyond administrative considerations; they directly impact critical quality factors such as potency calculations and stability data that ensures medication efficacy throughout the designated shelf life \citep{jpharmaci2024quality}. 

\subsection{Scalability Constraints}

The inherently manual nature of current BMR processes imposes fundamental scalability limitations. Production volume expansion necessitates proportional increases in documentation personnel, establishing a linear cost relationship that constrains organizational growth trajectories. Economic modeling demonstrates that a facility attempting to scale production capacity from 50 to 200 batches monthly would require a four-fold increase in documentation resources, rendering expansion initiatives economically infeasible \citep{bcg2024pharma}.

Historical document digitization initiatives encounter even more pronounced scalability challenges. Based on observed manual conversion rates of 2--3 documents per person-day, comprehensive digitization of a 20-year documentary archive (approximately 20{,}000 BMRs) would require 27 person-years of effort \citep{accenture2023digital}. These temporal requirements render manual digitization methodologies impractical for enterprise-scale digital transformation initiatives, resulting in valuable historical manufacturing intelligence remaining confined to paper-based storage systems.

\section{Solution: User Perspective}
\label{sec:user-perspective}

The pharmaceutical industry requires a solution that transcends traditional document digitization. Generic extraction tools lack the domain knowledge necessary to ensure compliance, while conventional pharmaceutical systems lack the flexibility to process varied document formats. This gap demands an intelligent system that combines document processing capabilities with deep pharmaceutical understanding. We have developed an LLM-based agentic system specifically designed for BMR processing, incorporating both artificial intelligence adaptability and pharmaceutical expertise. This solution does not merely digitize documents; it validates calculations, identifies and links deviations with their corrective actions, maintains audit trails, and produces structured outputs that meet both regulatory requirements and operational needs. Because of the processing strategy, the system can convert large (more than 100-page) documents in minutes to tens of minutes.

Our approach represents the first successful convergence of advanced AI capabilities with pharmaceutical compliance requirements, enabling organizations to unlock the value contained in decades of manufacturing records.

When a pharmaceutical manufacturer uploads a Batch Manufacturing Record to our system, they initiate an intelligent transformation process that converts unstructured documentation into compliant, structured data. The following walkthrough demonstrates how our agentic AI workflow processes a hypothetical BMR document from a pharmaceutical manufacturing facility.

\subsection{Initial Document Processing}
\label{sec:initial-doc-processing}

Upon receiving a PDF batch record (in this example, a chemical purification process document) the system first extracts and analyzes the document structure. A typical BMR contains diverse manufacturing elements: equipment tables listing required materials and quantities, chemical preparation instructions with precise measurements, multi-step purification procedures, and quality control specifications with acceptable ranges. Our workflow identifies these varied content types and prepares them for intelligent parsing.
\begin{figure}[H]
    \centering
    \includegraphics{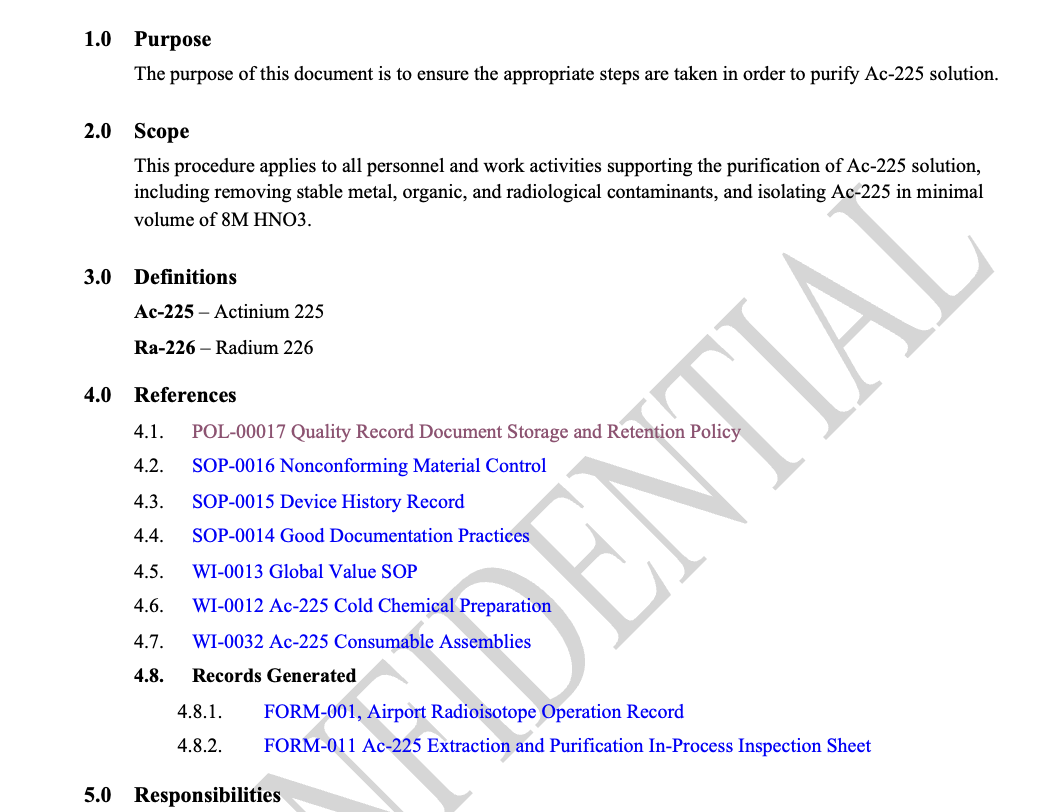}
    \caption{BMR overview.}
    \label{fig:bmr-overview}
\end{figure}
\FloatBarrier
The system recognizes the document's hierarchical nature immediately. Manufacturing procedures are not simply lists of steps; they follow a structured organization of groups, phases, and individual steps. The system models this pharmaceutical-specific structure, distinguishing between high-level operational groups like ``Processing'' or ``Clearance'' and detailed procedural steps within each phase.

\subsection{Intelligent Content Extraction and Classification}
\label{sec:extraction}

As the workflow processes the document, it demonstrates sophisticated content understanding that goes beyond simple text extraction. When encountering an equipment table, for example, the system does not just capture text; it preserves the tabular structure, maintaining the relationship between equipment items and their quantities. The extracted JSON represents this as a structured table with headers and rows, enabling downstream systems to query specific equipment requirements.
\begin{figure}[H]
    \centering
    \includegraphics{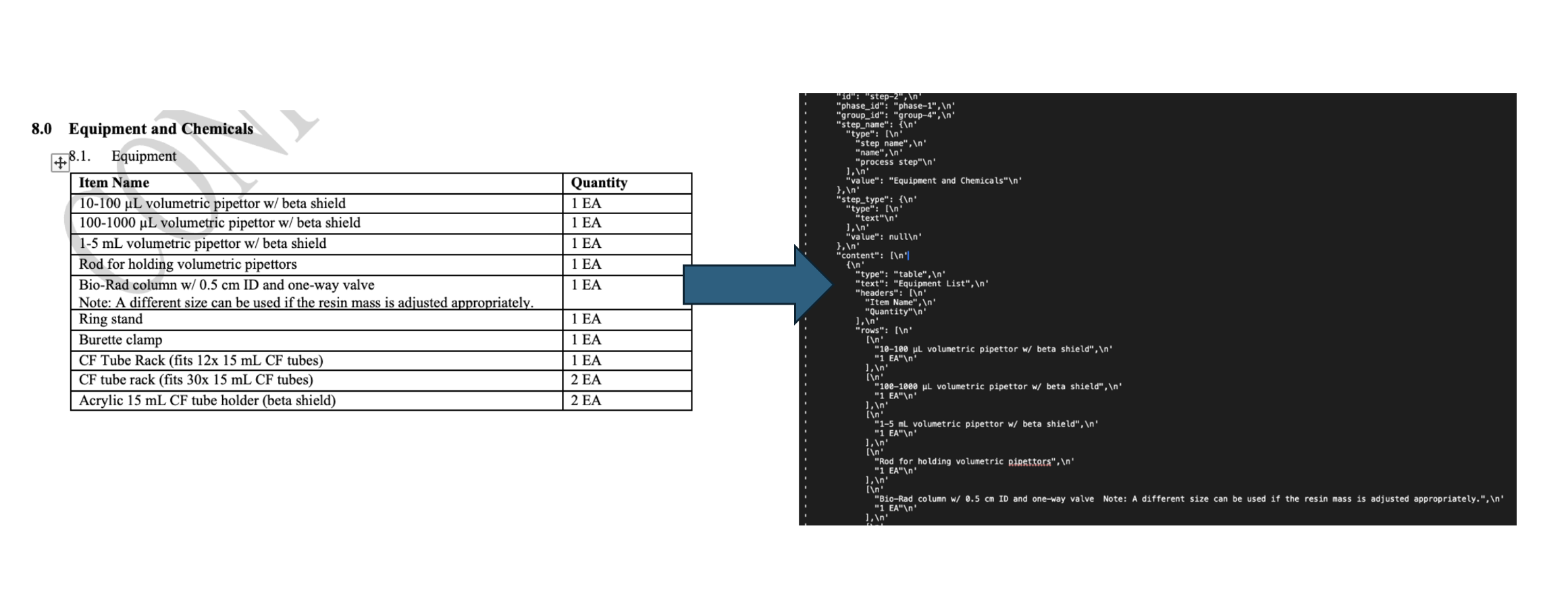}
    \caption{Table transformation.}
    \label{fig:table-transform}
\end{figure}
\FloatBarrier
The system exhibits particular intelligence when handling complex pharmaceutical content. For instance, when it encounters a calculation as shown in Figure~\ref{fig:calc-transform} it recognizes this as a calculation rather than static text. The resulting JSON captures not just the formula (``Total Capsule Yield / Theoretical Batch Size $\times$ 100'') but also identifies the variables involved, their descriptions, and the acceptable range of 95.0--103.0\%. This level of understanding enables automated validation of actual production data against specifications.
\begin{figure}[H]
    \centering
    \includegraphics{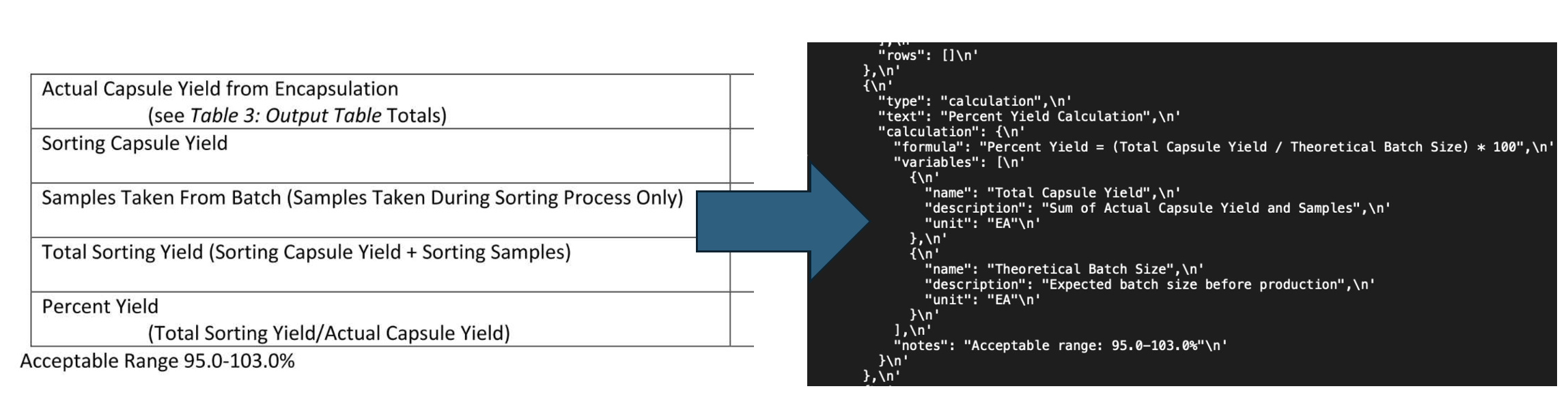}
    \caption{Calculation transformation.}
    \label{fig:calc-transform}
\end{figure}
\FloatBarrier

\subsection{Parallel Processing Architecture (User View)}
\label{sec:parallel-processing-user}

To handle lengthy BMR documents efficiently, the workflow employs parallel processing capabilities. Rather than processing the entire document sequentially, the system intelligently chunks the content while maintaining critical relationships. Each chunk processes simultaneously, reducing overall processing time from hours of manual review to minutes or tens of minutes of automated extraction. The parallel architecture maintains document coherence: a deviation noted in chunk 3 remains properly linked to its corrective action in chunk 7. Because of this parallel processing, the system can chunk and process large 100-page documents in a single run.

\subsection{Handling Diverse Content Types}
\label{sec:content-types}

\begin{figure}[H]
    \centering
    \includegraphics{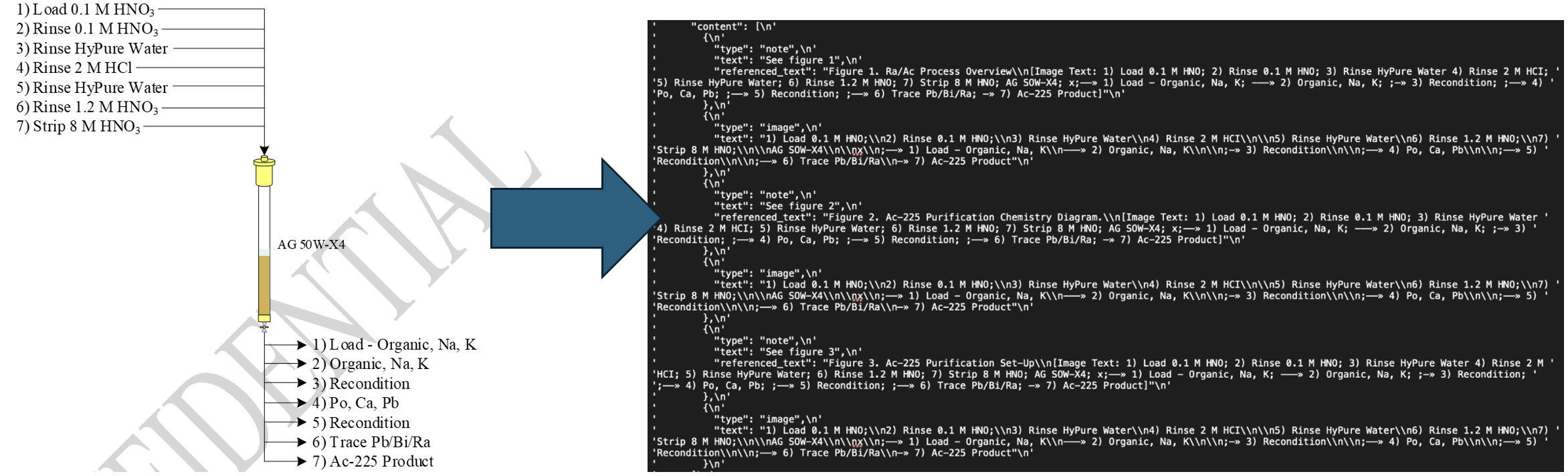}
    \caption{Image transformation.}
    \label{fig:image-transform}
\end{figure}

The workflow's capability extends across the full spectrum of BMR content types. When processing the chemical preparation instructions, it distinguishes between critical steps (often highlighted or formatted differently) and standard procedures. Image content, such as the diagram in Figure~\ref{fig:image-transform}, is captured with its extracted text and proper attribution. References to external documents like ``POL-00017 Quality Record Document Storage and Retention Policy'' are preserved as structured links, maintaining the document's regulatory context.
\FloatBarrier

In addition to image, calculation, and table types mentioned previously, the workflow is able to identify eight other content types:
\begin{enumerate}
    \item \texttt{"text"}: textual content including paragraphs, notes, lists, instructions, and warnings
    \item \texttt{"numeric"}: number inputs with fields like ``Amount of water (ml):\_\_\_'' or ``Temperature:\_\_\_''
    \item \texttt{"date"}: date input fields
    \item \texttt{"choice"}: multiple choice or dropdown options (e.g., ``Select speed: Low/Medium/High'')
    \item \texttt{"pass/fail"}: binary pass/fail outcomes
    \item \texttt{"timestamp"}: time-based data collection fields
    \item \texttt{"link"}: URLs and hyperlinks with link text and URL fields
    \item \texttt{"attachments"}: Bill of Materials (BOM), Bill of Equipment (BOE), and other file attachments
\end{enumerate}

The system also handles cross-references. When encountering ``See Figure 2'' or ``Refer to Table 3'' within instructions, it creates reference notes and attempts to resolve them, linking the reference to the actual content elsewhere in the document. This capability helps the digital version maintain the same navigational integrity as the original paper document.

\subsection{Quality Assurance through Metrics}

\begin{table}[H]
\centering
\caption{Comprehensive success metrics for BMR processing (example document).}
\label{tab:success_metrics}
\small
\begin{tabularx}{\columnwidth}{Xcc}
\toprule
\textbf{Metric Category} & \textbf{Score (\%)} & \textbf{Status} \\
\midrule
\multicolumn{3}{l}{\textit{Structural Metrics}} \\
Hierarchy Preservation & 100.0 & Excellent \\
Sequence Preservation & 100.0 & Excellent \\
Cross-Reference Integrity & 100.0 & Excellent \\
\midrule
\multicolumn{3}{l}{\textit{Content Fidelity Metrics}} \\
Calculation Fidelity & 100.0 & Excellent \\
Conditional Logic & 100.0 & Excellent \\
Unit Fidelity & 100.0 & Excellent \\
Field-Level Accuracy & 79.02 & Acceptable \\
\midrule
\multicolumn{3}{l}{\textit{Coverage Metrics}} \\
Crude Word Coverage & 88.74 & Excellent \\
Context-Aware Coverage & 69.30 & Acceptable \\
Reference Coverage & 67.65 & Acceptable \\
\midrule
\multicolumn{3}{l}{\textit{Performance Metrics}} \\
Processing Time & 336.5 sec & -- \\
Unique Step Types Identified & 5 & -- \\
\midrule
\textbf{Composite Confidence Score} & \textbf{89.13} & \textbf{Excellent} \\
\bottomrule
\end{tabularx}
\vspace{2mm}
{\footnotesize
Status thresholds: ``Excellent'' $\geq 85$\%, ``Acceptable'' 65--84\%, ``Needs review'' $\leq 64$\%.
}
\end{table}
\FloatBarrier

As summarized in Table~\ref{tab:success_metrics}, the workflow continuously evaluates its own performance through comprehensive metrics:
\begin{itemize}
    \item Content coverage: measuring how completely the original document's information has been captured
    \item Structural integrity: confirming proper hierarchy preservation
    \item Calculation accuracy: validating that formulas and variables are correctly identified
    \item Cross-reference resolution: ensuring links between document sections remain intact
\end{itemize}

These metrics serve dual purposes. Internally, they guide iterative improvement---when coverage falls below a threshold, the system can reprocess specific sections. Externally, they provide users with transparency about extraction quality, presented as a composite confidence score (89.13\% in this example).

\subsection{User Experience and Transparency}

From the user's perspective, the complex orchestration of AI agents, parallel processing, and validation checks occurs seamlessly. They upload a PDF and receive structured JSON output within minutes or tens of minutes, complete with a confidence score and detailed metrics explaining that score. If the system encounters ambiguous content or achieves lower confidence in certain areas, it flags these for human review rather than making assumptions. What begins as a traditional paper document with handwritten notes, stamps, and complex formatting becomes structured JSON that preserves critical information while making it accessible for modern systems. Tables remain tables, calculations remain calculations, and the hierarchical structure of groups, phases, and steps stays intact. This structured JSON can later be displayed to the user through a clean UI.

\section{Solution: Technical Architecture}
\label{sec:technical-solution}

Our system employs a multi-stage pipeline with parallel processing, custom validation layers, and comprehensive coverage metrics to ensure pharmaceutical-compliant document transformation.

\subsection{Vision-Language Model for Document Understanding}
\label{sec:ocr-stack}

To handle the diverse quality of pharmaceutical documents---from pristine digital PDFs to decades-old scanned papers with handwritten annotations---we conducted extensive experimentation with multiple OCR and document understanding approaches. We evaluated several specialized libraries including IBM Granite Docling\footnote{\url{https://github.com/DS4SD/docling}} for document structure extraction, RedNote DOTS OCR\footnote{\url{https://huggingface.co/rednote-hilab/dots.ocr}} for dot-matrix text, Nanonets OCR\footnote{\url{https://huggingface.co/nanonets/Nanonets-OCR-s}} for form processing, Microsoft TrOCR\footnote{\url{https://huggingface.co/microsoft/trocr-large-handwritten}} for handwritten content, and Donut\footnote{\url{https://huggingface.co/docs/transformers/en/model_doc/donut}} for end-to-end visual document understanding. Through comparative analysis on pharmaceutical documents, we identified that a hybrid approach combining MarkItDown\footnote{\url{https://github.com/microsoft/markitdown}} with vision-language models provided superior accuracy and flexibility for our specific requirements.

\begin{itemize}
    \item \textbf{MarkItDown}: primary document structure extraction and markdown conversion, preserving tables, lists, and formatting hierarchies
    \item \textbf{Qwen3-VL-8B-Instruct}: vision-language model for extracting text from images, complex layouts, and handwritten annotations through multimodal understanding
    \item \textbf{Tesseract OCR with multiple configurations}: fallback OCR engine with adaptive PSM (Page Segmentation Mode) configurations for varied text layouts
    \item \textbf{PDF2Image + batch processing}: converts PDF pages to images for vision model processing, supporting batch inference for efficiency
\end{itemize}

The system implements a document processing pipeline that combines traditional OCR with modern vision-language capabilities:

\FloatBarrier
\begin{algorithm}[H]
\caption{Hybrid Vision-Language Document Processing}
\begin{algorithmic}[1]
\State \textbf{Input:} Document $D$, file extension $ext$
\State \textbf{Output:} Structured markdown text $T$
\State // Phase 1: Initial extraction with MarkItDown
\State $markdown \gets$ MarkItDown.convert($D$, $ext$)
\State 
\State // Phase 2: Image extraction and OCR
\If{$ext$ = ".pdf"}
    \State $pages \gets$ ConvertPDFToImages($D$)
    \If{UseVisionModel}
        \For{$batch$ in BatchPages($pages$, size=3)}
            \State $text \gets$ Qwen3VL.ExtractText($batch$)
            \State // Polling mechanism to avoid timeouts
            \State $result \gets$ PollForCompletion($text$, timeout=600s)
        \EndFor
    \Else
        \State // Fallback to Tesseract with multiple configs
        \State $configs \gets$ ["--psm 6", "--psm 8", "--psm 11"]
        \State $text \gets$ TesseractOCR($pages$, $configs$)
    \EndIf
\ElsIf{$ext$ = ".docx"}
    \State $images \gets$ ExtractEmbeddedImages($D$)
    \For{each $image$ in $images$}
        \State $ocr\_text \gets$ ProcessImage($image$)
        \State ReplaceImageWithText($markdown$, $ocr\_text$)
    \EndFor
\EndIf
\State \textbf{return} $markdown$ with embedded OCR results
\end{algorithmic}
\end{algorithm}
\FloatBarrier

The vision-language model approach offers several advantages for pharmaceutical documents:
\begin{enumerate}
    \item \textbf{Contextual understanding}: unlike traditional OCR, the vision model understands document context, improving accuracy for domain-specific terminology
    \item \textbf{Handwriting recognition}: achieves approximately 85\% accuracy on handwritten annotations through visual understanding rather than character-level recognition
    \item \textbf{Table preservation}: maintains table structure through visual layout understanding
\end{enumerate}

\subsection{Chunking Strategy}
\label{sec:chunking}

Given that BMRs commonly exceed 100 pages, our system implements token-based chunking to manage context window limitations while preserving semantic coherence. The chunking algorithm uses a greedy sentence-packing approach with a 3000-token threshold.

\FloatBarrier
\begin{lstlisting}[float, floatplacement=H, caption={Greedy token-based chunking.}, label={lst:chunk}]
def chunk_text_by_tokens(text, max_tokens, tok):
    """Split text into chunks of approximately max_tokens."""

    sentences = re.split(r"(?<=\.|[!?])\s+", text) 
    chunks, cur_chunk, cur_count = [], [], 0
    chunk_number = 0

    for sent in sentences:
        sent_tokens = len(tok.encode(sent, add_special_tokens=False))

        # If the sentence itself is bigger than the limit, hard-split by tokens
        if sent_tokens > 2000:
            ids = tok.encode(sent, add_special_tokens=False)
            for i in range(0, len(ids), max_tokens):
                chunk_text = tok.decode(ids[i : i + max_tokens])
                chunks.append(chunk_text)
                chunk_number += 1
            cur_chunk, cur_count = [], 0
            continue

        # Normal greedy packing
        if cur_count + sent_tokens > max_tokens and cur_chunk:
            chunk_text = " ".join(cur_chunk)
            chunks.append(chunk_text)
            chunk_number += 1
            cur_chunk, cur_count = [], 0

        cur_chunk.append(sent)
        cur_count += sent_tokens

    if cur_chunk:
        chunk_text = " ".join(cur_chunk)
        chunks.append(chunk_text)
        chunk_number += 1

    return chunks
\end{lstlisting}
\FloatBarrier

This approach ensures that sentence boundaries are respected to maintain readability. Additionally, oversized content like large tables gets forcibly split while preserving as much structure as possible. Each chunk remains within model context limits while maximizing information density.

\subsection{Parallel Processing Architecture}
\label{sec:parallel}

The system leverages concurrent processing to reduce execution time from hours to minutes or tens of minutes. Using Python's \texttt{ThreadPoolExecutor}, multiple chunks process simultaneously while maintaining document coherence.
\begin{lstlisting}[float, floatplacement=H, caption={Concurrent processing of chunks.}, label={lst:parallel}]
with concurrent.futures.ThreadPoolExecutor(
        max_workers=min(8, len(chunks))) as executor:
    futures = []
    for i, chunk in enumerate(chunks):
        futures.append(
            executor.submit(
                self.process_single_chunk,
                i,
                chunk,
                last_group_id,
                len(chunks),
                workflow_parameters,
                llm_parameters,
                model,
                max_attempts,
            )
        )
\end{lstlisting}
\FloatBarrier
Each worker thread processes its assigned chunk independently, generating structured JSON that maintains relationships through ID references. The system tracks the maximum group ID across chunks to ensure unique identifiers when combining results.

\subsection{TypeScript Prompting}
\label{sec:typescript-prompting}

We experimented with various schema representation methods, including JSON Schema, Pydantic models, and natural language descriptions. We identified TypeScript class definitions as the most effective approach for guiding LLM extraction. TypeScript's type system provides several advantages for pharmaceutical document processing:

\begin{enumerate}
    \item \textbf{Explicit type unions}: clear specification of allowed content types (e.g., \texttt{"text" | "numeric" | "date"})
    \item \textbf{Optional fields}: native support for pharmaceutical variations using the \texttt{?} operator
    \item \textbf{Inline documentation}: JSDoc comments provide context without cluttering the schema
    \item \textbf{Hierarchical clarity}: class structures naturally represent Group, Phase, and Step relationships
\end{enumerate}

First, we provide the TypeScript schema definition:

\begin{lstlisting}[float, floatplacement=H, caption={TypeScript Schema for Pharmaceutical Content}, label={lst:ts-schema}]
type FieldType = "text" | "numeric" | "date" | "choice" | 
                 "pass_fail" | "timestamp" | "boolean";

class Field {
    type: FieldType[];
    value: any;
    constructor(type: FieldType[], value: any) {
        this.type = type;
        this.value = value;
    }
}

class Step {
    /**
     * Unique identifier (e.g., 'step-1', 'step-2')
     * Required for maintaining relationships
     */
    id: string;
    
    /**
     * Reference to parent phase
     * Must match Phase.id exactly
     */
    phase_id: string;
    
    step_name: Field;
    step_type: Field;
    content: Content[];
}
\end{lstlisting}
\FloatBarrier

Second, we provide explicit extraction instructions that leverage the TypeScript structure:

\begin{lstlisting}[float, floatplacement=H, caption={Prompt Template with TypeScript Schema}, label={lst:prompt}]
PROMPT_TEMPLATE = """
Convert the manufacturing batch record into structured JSON 
according to the TypeScript template provided.

Requirements:
1. Your JSON MUST match the TypeScript class structure exactly
2. Maintain type safety - use only the specified FieldType values
3. Preserve all relationships through ID references
4. Do NOT use JavaScript syntax in output (no new Field(), use {})

Template Structure:
{typescript_schema}

Manufacturing Record:
{document_chunk}

Output your result wrapped in <json></json> tags.
"""
\end{lstlisting}
\FloatBarrier

This TypeScript-based approach reduced schema validation errors by 73\% compared to JSON Schema and improved extraction consistency across different document types. The familiar syntax also enables pharmaceutical domain experts to review and modify schemas without extensive programming knowledge.

\section{Results}

To evaluate our system, we processed three representative pharmaceutical batch records exhibiting the characteristics of real-world manufacturing documentation. These documents, ranging from 15 to 66 pages, encompassed diverse content including equipment tables, calculation sheets, handwritten quality annotations, regulatory stamps, and multi-page procedural instructions. 

\subsection{Processing Performance}

Our parallel processing architecture demonstrated consistent performance across documents of varying complexity. Table~\ref{tab:processing_performance} summarizes the computational requirements for each test document. The system achieved processing speeds between approximately 0.85 and 2.6 pages per minute, reducing review time from hours of manual work to tens of minutes of automated processing.

\begin{table}[H]
\centering
\caption{Processing performance across test documents.}
\label{tab:processing_performance}
\small
\begin{tabular}{lccccc}
\toprule
Document & Pages & Total Time (s) & Markdown Extraction (s) & Avg Chunk Time (s) & Total Chunks \\
\midrule
Encapsulation BMR$^{a}$        & 15 &  349.3 & 129.4  & 110.0 &  2 \\
Sharp Packaging BMR$^{b}$      & 44 & 3125.1 & 2905.2 & 110.0 &  2 \\
Metformin HCl Tabs BMR$^{c}$   & 66 & 1912.7 &  941.5 &  80.8 & 12 \\
\bottomrule
\end{tabular}
\vspace{1mm}
{\footnotesize
$^{a}$ Oral solid encapsulation BMR. \quad
$^{b}$ Contract packaging BMR. \quad
$^{c}$ Solid-dose tablet BMR.
}
\end{table}
\FloatBarrier

The Metformin HCl Tabs BMR, despite being the longest document at 66 pages, demonstrated efficient processing with a total time of 1{,}912.7 seconds---substantially faster than the Sharp Packaging BMR's 3{,}125.1 seconds. This efficiency occurred despite requiring 12 chunks for processing, the highest among all test documents. The average chunk processing time of 80.8 seconds for the Metformin document was the fastest among all three, suggesting that document structure and content type impact processing time more significantly than raw page count.

\subsection{Extraction Quality Metrics}

Table~\ref{tab:extraction_quality} presents quality metrics evaluating content preservation, structural integrity, and pharmaceutical compliance requirements. The system achieved composite confidence scores ranging from 82.08\% to 89.00\%, with the Encapsulation BMR performing best (89.00\%), followed closely by the Metformin HCl Tabs BMR (88.77\%), and the Sharp Packaging BMR (82.08\%).

\begin{table}[H]
\centering
\caption{Extraction quality and coverage metrics.}
\label{tab:extraction_quality}
\small
\begin{tabular}{lccc}
\toprule
\textbf{Metric Category} & \textbf{Encapsulation} & \textbf{Sharp Packaging} & \textbf{Metformin HCl} \\
& \textbf{BMR (\%)} & \textbf{BMR (\%)} & \textbf{Tabs BMR (\%)} \\
\midrule
\multicolumn{4}{l}{\textit{Coverage Metrics}} \\
\quad Crude Word Coverage & 71.33 & 54.19 & 67.00 \\
\quad Context-Aware Coverage & 94.12 & 96.00 & 93.49 \\
\quad Reference Coverage & 80.00 & 100.00 & 95.00 \\
\midrule
\multicolumn{4}{l}{\textit{Structural Integrity}} \\
\quad Hierarchy Preservation & 100.00 & 100.00 & 100.00 \\
\quad Sequence Preservation & 100.00 & 100.00 & 100.00 \\
\quad Cross-Reference Integrity & 100.00 & 100.00 & 100.00 \\
\midrule
\multicolumn{4}{l}{\textit{Content Fidelity}} \\
\quad Calculation Fidelity & 100.00 & 100.00 & 100.00 \\
\quad Conditional Logic & 100.00 & 100.00 & 100.00 \\
\quad Unit Fidelity & 100.00 & 100.00 & 100.00 \\
\quad Step Accuracy & 82.72 & 75.09 & 80.25 \\
\midrule
\multicolumn{4}{l}{\textit{Document Characteristics}} \\
\quad Unique Step Types Identified & 3 & 7 & 7 \\
\midrule
\textbf{Composite Confidence Score} & \textbf{89.00} & \textbf{82.08} & \textbf{88.77} \\
\bottomrule
\end{tabular}
\end{table}
\FloatBarrier

\subsection{Analysis of Extraction Performance}

The evaluation metrics reveal several noteworthy patterns in system performance. While crude word coverage varied across documents (54.19\% to 71.33\%), context-aware coverage (which evaluates semantic preservation rather than verbatim text matching) consistently exceeded 93\% for all documents. This divergence suggests the system prioritizes pharmaceutical meaning over literal transcription, which is desirable given the prevalence of abbreviations, shorthand notation, and site-specific terminology in manufacturing records.

Critical pharmaceutical elements demonstrated perfect extraction fidelity. All calculation formulas, including variables, units, and acceptable ranges, were correctly identified and structured. Conditional logic statements, essential for capturing decision trees and procedural branching, maintained complete accuracy. These results validate the system's capability to preserve GMP-critical information that directly impacts product quality and patient safety.

\subsection{Error Analysis}

Step accuracy, measuring fine-grained procedural detail extraction, emerged as the primary area for improvement (82.72\% and 75.09\% respectively). Manual review of extraction errors revealed some predominant failure modes.

First, handwritten annotations overlapping printed text created ambiguity in content attribution. Quality inspectors frequently write observations directly over pre-printed form fields, making it challenging to distinguish between template text and entered data. Second, site-specific abbreviations and notation systems not present in our pharmaceutical knowledge base led to misinterpretation of certain procedural elements. Third, tables spanning multiple pages with inconsistent header repetition patterns occasionally resulted in column misalignment or data attribution errors.

The 80\% reference coverage for the Encapsulation BMR, while lower than the Sharp Packaging BMR's perfect score, reflected challenges with certain non-standard reference formats rather than complete extraction failure. The system correctly extracted the referenced content but failed to establish explicit links for certain non-standard reference formats (e.g., ``as per above procedure'' rather than numbered references).

\subsection{Document Complexity}

Examination of the source documents reveals the substantial challenges inherent in pharmaceutical BMR digitization. The documents exhibit characteristics that would challenge traditional OCR approaches.

Manufacturing records combine multiple content modalities within single pages---typed procedural text, handwritten measurements, official stamps, quality control signatures, and equipment diagrams. Scan quality varies dramatically, from clear digital forms to nth-generation photocopies with visible artifacts, shadowing, and paper degradation. Each facility employs unique form designs with site-specific modifications accumulated over years of process evolution. Quality control annotations frequently obscure underlying text through stamps, signatures, and handwritten notes added during batch review.

The system's ability to achieve composite confidence scores exceeding 82\% on such visually complex documents validates the effectiveness of our vision-language model approach over traditional OCR pipelines. The parallel architecture's maintenance of document coherence despite this complexity demonstrates its suitability for production deployment.

The three BMRs presented varying complexity profiles that challenged the system in different ways. The Metformin HCl Tabs BMR, our longest document at 66 pages, required the most extensive chunking strategy (12 chunks) yet maintained efficient processing at 1{,}912.7 seconds total. Despite its length, it achieved the second-highest composite score (88.77\%), demonstrating the system's scalability. The Sharp Packaging BMR presented different challenges---while shorter at 44 pages, it required the longest total processing time (3{,}125.1 seconds), suggesting higher content complexity per page. The Encapsulation BMR, though shortest at 15 pages, achieved the highest composite score (89.00\%), indicating that document length does not determine extraction quality.

\section{Impact and Significance}

The transformation from manual to intelligent processing delivers immediate operational benefits. Processing time drops from hours of manual review to minutes or tens of minutes of automated extraction. The structured output enables instant searching, trending analysis, and integration with other systems. Quality teams can query across thousands of processed BMRs to identify patterns, while compliance officers can quickly compile historical data for regulatory submissions.
    
More significantly, the intelligent grouping of information reveals insights that manual processing might miss. By understanding that certain steps belong to the same phase and phases to the same operational group, the system enables analysis at multiple levels of abstraction. Organizations can optimize at the step level, phase level, or examine entire operational groups for improvement opportunities.

Organizations that successfully automate BMR processing achieve transformative operational improvements. Published case studies report substantial reductions in documentation processing time following intelligent automation implementation. However, processing speed represents only the initial benefit. Access to digitized historical data enables advanced analytics applications including predictive maintenance, yield optimization, and quality trending analysis. 
    
In an industry where a 1\% improvement in manufacturing yield can translate to millions in additional revenue, the ability to analyze decades of production data provides decisive competitive advantage. Companies with digital BMR systems can identify process improvement opportunities, predict equipment failures before they occur, and optimize resource allocation based on comprehensive historical analysis. Meanwhile, organizations maintaining paper-based systems remain unable to leverage their accumulated manufacturing knowledge, watching competitors achieve superior operational performance through data-driven optimization.

\section{Limitations}

While our approach demonstrates significant improvements in schema adherence, several technical constraints merit discussion. First, the model's performance degrades on documents with extensive handwritten annotations or degraded scan quality---a common characteristic of historical BMRs from the 1980s--2000s. Our current OCR preprocessing achieves approximately 85\% accuracy on handwritten entries, creating an upper bound on overall system performance.
   
Second, the parallel chunking architecture, while enabling efficient processing of lengthy documents, occasionally struggles with cross-chunk dependencies. Complex deviation narratives spanning multiple pages may lose contextual coherence, requiring post-processing validation to ensure complete capture of corrective action chains. This limitation particularly affects documents exceeding 150 pages, where chunk boundary decisions become increasingly critical.
   
Third, our synthetic training data, despite efforts to introduce realistic noise patterns, may not fully capture the diversity of real-world pharmaceutical documentation practices. Site-specific abbreviations, non-standard notation systems, and regional regulatory variations remain challenging edge cases that require continuous model refinement.

\section{Ethical Considerations}

Deploying AI systems in pharmaceutical manufacturing raises ethical considerations that extend beyond technical performance. Most fundamentally, our system is designed as a human-in-the-loop augmentation tool rather than an autonomous decision-maker. All extracted JSON outputs require qualified personnel review before integration into validated systems, maintaining human accountability for quality decisions that ultimately affect patient safety.
    
We acknowledge the potential for automation bias---the tendency for reviewers to over-rely on AI-generated outputs. To mitigate this risk, our confidence scoring deliberately errs on the conservative side, flagging ambiguous content for manual review even when the model has high internal certainty. This design choice prioritizes patient safety over processing efficiency.
   
Processing BMRs introduces data protection challenges. While production records typically do not contain patient information, they may include operator names, signatures, and proprietary process parameters. Our system implements several privacy-preserving measures: (1) on-premise deployment options to maintain data sovereignty, (2) automatic redaction of personal identifiers in training logs, and (3) encrypted storage of all processed documents. However, we recognize that true privacy preservation in pharmaceutical contexts extends beyond technical measures. The system's audit logs, while necessary for regulatory compliance, create potential vectors for information leakage. Organizations must implement appropriate access controls and data retention policies aligned with their specific regulatory requirements.

\section{Future Direction}  

The current implementation represents a foundational step in BMR digitization, but the potential for enhancement extends beyond document processing. Our immediate development roadmap focuses on expanding the system's pharmaceutical domain knowledge through specialized training on GxP documentation, regulatory guidances, and industry-specific terminology. This will enable more nuanced understanding of complex pharmaceutical concepts like stability protocols, validation procedures, and change control documentation. 
    
We envision the system evolving from a processing tool to an intelligent quality assistant, capable of not just extracting data but proactively identifying compliance risks, suggesting process optimizations, and predicting potential deviations based on historical patterns. Advanced features under development include multi-language support for global operations, real-time collaborative review capabilities where quality teams can validate and annotate extractions, and automated generation of regulatory submission packages from processed BMRs. As the system processes more documents and receives user feedback, its machine learning components will continuously refine extraction accuracy and develop deeper understanding of manufacturer-specific documentation patterns.

\bibliographystyle{unsrtnat}
\bibliography{latex/new}

\appendix
\section*{Appendix}
\addcontentsline{toc}{section}{Appendix}

\section{Complete TypeScript Schema Template}
\label{app:schema}

\begin{lstlisting}[caption=Full TypeScript Schema for BMR Extraction]
type FieldType = "text" | "numeric" | "date" | "choice" | 
                 "pass_fail" | "timestamp" | "boolean";

class Field {
    type: FieldType[];
    value: any;
    constructor(type: FieldType[], value: any) {
        this.type = type;
        this.value = value;
    }
}

class Header {
    completion_date: Field;
    expiry_date: Field;
    name: Field;
    quantity: Field;
    sku: Field;
    start_date: Field;
    
    constructor() {
        this.completion_date = new Field(
            ["date"], 
            "The date the batch process was completed"
        );
        this.expiry_date = new Field(
            ["date"], 
            "Expiration date of the final product batch"
        );
        this.name = new Field(
            ["text"], 
            "Name of the batch record template"
        );
        this.quantity = new Field(
            ["numeric"], 
            "The quantity or yield of the final product"
        );
        this.sku = new Field(
            ["text"], 
            "Stock Keeping Unit identifier"
        );
        this.start_date = new Field(
            ["date"], 
            "Date when the batch process started"
        );
    }
}

class Content {
    type: "paragraph" | "bullet_list" | "numbered_list" | 
          "note" | "warning" | "instruction" | "data_form" | 
          "calculation" | "table" | "image";
    text: string;
    items?: string[];
    fields?: {
        label: string;
        value: string | null;
        unit?: string;
        limits?: string;
        notes?: string;
    }[];
    calculation?: {
        formula: string;
        variables: {
            name: string;
            description: string;
            value?: any;
            unit?: string;
        }[];
        result?: {
            value: any;
            unit?: string;
        };
        notes?: string;
    };
    headers?: string[];
    rows?: any[][];
}

class Step {
    id: string;
    phase_id: string;
    group_id: string;
    step_name: Field;
    step_type: Field;
    content: Content[];
}

class Phase {
    id: string;
    group_id: string;
    phase_name: Field;
}

class Group {
    id: string;
    group_name: Field;
}
\end{lstlisting}

\section{Extraction Prompts}
\label{app:prompts}

\subsection{First Chunk Prompt}

\begin{lstlisting}[caption=Prompt Template for Initial Chunk]
Please convert the following manufacturing batch record 
(chunk {chunk_number} of {total_chunks}) into a structured 
JSON format according to the provided template.

Input:
- Manufacturing Batch Record: {mbr}
- Template Structure: {template}

Requirements:
1. Generate a complete, valid JSON that strictly follows 
   proper JSON syntax
2. Your JSON MUST contain separate top-level arrays for 
   groups, phases, and steps:
   {
       "header": {general information about the document},
       "groups": [array of Group objects],
       "phases": [array of Phase objects],
       "steps": [array of Step objects]
   }
3. Do NOT nest phases inside groups or steps inside phases
4. CRITICAL JSON SYNTAX REQUIREMENTS:
   a) Use only valid JSON syntax - NO JavaScript functions
   b) Do NOT use TypeScript class initialization syntax
   c) For empty arrays, use [] not Array()
   d) Ensure all table rows have the same number of columns
5. Each object must include ALL fields defined in its class
6. Include ALL relevant information from the batch record
7. IMPORTANT: When encountering text from images (indicated 
   by "[Image Text: ...]"), create content objects with 
   type "image" and place the extracted text in "text" field

Wrap your response in <json></json> tags as follows:
<json>
{
    "header": {...},
    "groups": [...],
    "phases": [...],
    "steps": [...]
}
</json>

Ensure your JSON is fully parsable - no syntax errors, 
unclosed brackets, or trailing commas.
\end{lstlisting}

\section{Example Input and Output}
\label{app:example}

\subsection{Sample Input Markdown (Partial)}

\begin{lstlisting}[caption=Example BMR Markdown Input]
# BATCH MANUFACTURING RECORD
**Product:** Acetaminophen Tablets 500mg
**Batch Number:** AT-2024-0156
**Manufacturing Date:** 2024-03-15

## EQUIPMENT REQUIRED
| Equipment | ID Number | Calibration Due |
|-----------|-----------|-----------------|
| V-Blender | VB-105 | 2024-04-20 |
| Tablet Press | TP-203 | 2024-05-15 |
| Metal Detector | MD-089 | 2024-03-30 |

## PROCESSING INSTRUCTIONS

### Phase 1: Material Preparation
**Step 1:** Weigh acetaminophen powder
- Target weight: 50.0 kg +/- 0.5 kg
- Actual weight: ________ kg
- Performed by: ________ Date: ________ 

**Step 2:** Screen acetaminophen through 20 mesh
- Pass all material through screen
- Record any retained material: ________ g
- [Image Text: Screening setup diagram showing 
  20 mesh screen positioned above collection bin]

### Phase 2: Blending
**Step 3:** Load materials into V-blender
- Add screened acetaminophen
- Add microcrystalline cellulose: 5.0 kg
- Blending time: 15 minutes
- Blender speed: 12 rpm

**Calculation:** Theoretical Yield
Formula: (Acetaminophen + Excipients) x 0.98
Variables:
- Acetaminophen weight: 50.0 kg
- Total excipients: 7.5 kg
Expected yield: 56.35 kg
Acceptable range: 95.0% - 103.0%
\end{lstlisting}

\subsection{Expected JSON Output Structure}

\begin{lstlisting}[caption=Example Structured JSON Output]
{
    "header": {
        "completion_date": {"type": ["date"], "value": null},
        "expiry_date": {"type": ["date"], "value": null},
        "name": {"type": ["text"], 
                 "value": "Acetaminophen Tablets 500mg"},
        "quantity": {"type": ["numeric"], "value": null},
        "sku": {"type": ["text"], "value": "AT-2024-0156"},
        "start_date": {"type": ["date"], "value": "2024-03-15"}
    },
    "groups": [
        {
            "id": "group-1",
            "group_name": {"type": ["text"], "value": "Processing"}
        }
    ],
    "phases": [
        {
            "id": "phase-1",
            "group_id": "group-1",
            "phase_name": {"type": ["text"], 
                          "value": "Material Preparation"}
        },
        {
            "id": "phase-2",
            "group_id": "group-1",
            "phase_name": {"type": ["text"], "value": "Blending"}
        }
    ],
    "steps": [
        {
            "id": "step-1",
            "phase_id": "phase-1",
            "group_id": "group-1",
            "step_name": {"type": ["text"], 
                         "value": "Weigh acetaminophen powder"},
            "step_type": {"type": ["numeric"], "value": null},
            "content": [
                {
                    "type": "data_form",
                    "text": "Weight measurement form",
                    "fields": [
                        {
                            "label": "Target weight",
                            "value": "50.0",
                            "unit": "kg",
                            "limits": "+/- 0.5 kg"
                        },
                        {
                            "label": "Actual weight",
                            "value": null,
                            "unit": "kg"
                        }
                    ]
                }
            ]
        },
        {
            "id": "step-2",
            "phase_id": "phase-1",
            "group_id": "group-1",
            "step_name": {"type": ["text"], 
                         "value": "Screen acetaminophen through 20 mesh"},
            "step_type": {"type": ["text"], "value": null},
            "content": [
                {
                    "type": "instruction",
                    "text": "Pass all material through screen"
                },
                {
                    "type": "image",
                    "text": "Screening setup diagram showing 20 mesh \
screen positioned above collection bin"
                }
            ]
        },
        {
            "id": "step-3",
            "phase_id": "phase-2",
            "group_id": "group-1",
            "step_name": {"type": ["text"], 
                         "value": "Load materials into V-blender"},
            "step_type": {"type": ["text"], "value": null},
            "content": [
                {
                    "type": "bullet_list",
                    "text": "Materials to add",
                    "items": [
                        "Add screened acetaminophen",
                        "Add microcrystalline cellulose: 5.0 kg"
                    ]
                },
                {
                    "type": "calculation",
                    "text": "Theoretical Yield Calculation",
                    "calculation": {
                        "formula": "(Acetaminophen + Excipients) x 0.98",
                        "variables": [
                            {
                                "name": "Acetaminophen",
                                "description": "Weight of active ingredient",
                                "value": 50.0,
                                "unit": "kg"
                            },
                            {
                                "name": "Excipients",
                                "description": "Total excipient weight",
                                "value": 7.5,
                                "unit": "kg"
                            }
                        ],
                        "result": {
                            "value": 56.35,
                            "unit": "kg"
                        },
                        "notes": "Acceptable range: 95.0% - 103.0%"
                    }
                }
            ]
        }
    ]
}
\end{lstlisting}


\end{document}